\documentclass[footinbib,a4paper,aps,prl,reprint,twocolumn,preprintnumbers,amsmath,amssymb,10pt, superscriptaddress]{revtex4-1}
\usepackage{amsmath}
\usepackage{verbatim}
\usepackage{graphicx}
\usepackage{color}
\usepackage{tikz}
\usepackage[colorlinks=true,citecolor=blue,linkcolor=blue,urlcolor=blue]{hyperref}
\usepackage{braket, enumitem,}
\usepackage[normalem]{ulem}
\usepackage{upgreek}

\begin{document}

\title{Phase diagram of 1+1D Abelian-Higgs model and its critical point}

\author{Titas Chanda} \email{tchanda@ictp.it}
\affiliation{The Abdus Salam International Centre for Theoretical Physics (ICTP), Strada Costiera 11, 34151 Trieste, Italy}
\affiliation{Instytut Fizyki Teoretycznej, Uniwersytet Jagiello{\'n}ski, \L{}ojasiewicza 11, 30-348 Krak{\'o}w, Poland}

\author{Maciej Lewenstein}
\affiliation{ICFO-Institut de Ci\`encies Fot\`oniques, The Barcelona Institute of Science and Technology, Av. Carl Friedrich
Gauss 3, 08860 Barcelona, Spain}
\affiliation{ICREA, Passeig Lluis Companys 23, 08010 Barcelona, Spain}

\author{Jakub Zakrzewski}
\affiliation{Instytut Fizyki Teoretycznej, Uniwersytet Jagiello{\'n}ski, \L{}ojasiewicza 11, 30-348 Krak{\'o}w, Poland}
\affiliation{Mark Kac Complex Systems Research Center, Jagiellonian University in Krakow, \L{}ojasiewicza 11, 30-348 Krak\'ow,
 Poland. }

\author{Luca Tagliacozzo}
\affiliation{Instituto de Física Fundamental IFF-CSIC, Calle Serrano 113b, Madrid 28006, Spain}
\affiliation{Departament de F\'{\i}sica Qu\`antica i Astrof\'{\i}sica and Institut de Ci\`encies del Cosmos (ICCUB), Universitat de Barcelona,  Mart\'{\i} i Franqu\`es 1, 08028 Barcelona, Catalonia, Spain}



\begin{abstract}
We  determine the phase diagram of the Abelian-Higgs model in one spatial dimension and time (1+1D) on a lattice. We identify a line of first order phase transitions separating the Higgs region from the confined one. This line terminates in a  quantum critical point above which  the two regions are connected by a smooth crossover. We analyze the critical point and find compelling evidences for its description  as the product of two non-interacting systems, a massless free fermion and a massless free boson. However, we find also some surprizing results  that cannot be explained by our simple picture, suggesting this newly discovered critical point to be an unusual one.
\end{abstract}

\maketitle

\paragraph{Introduction.}
Gauge theories in 1+1 dimensions  (1D in space and time) are ideal playgrounds to characterize the effects of strong-coupling between  matter and  gauge fields. Many of the non-perturbative aspects of 3+1 dimensional gauge theories relevant to our understanding of  particle physics, such as quark confinement and chiral symmetry breaking, have a 1+1D analogue. Furthermore, 1+1D field theories can often be treated analytically \cite{borgs_phase_1987, brydges_diamagnetic_1979} providing important insights to the physics of 1+1D systems relevant also to condensed matter physics.

In this work, we study the lattice version of a relativistic bosonic field that interacts with a photonic field  \cite{chanda20}, the bosonic version of the Schwinger model  \cite{schwinger_pr_1951, schwinger_pr_1962, schwinger_pr_1962_2, coleman_aop_1976}. In contrast to (polarized) fermions, bosons can have contact interactions that are described by the well known  Abelian-Higgs model (AHM)  in 1+1D (AHM$_2$) \cite{anderson_gauge_63, englert_broken_64, higgs_broken_64, Guralnik_conservation_64, peskin_book}.

In  AHM${}_2$, a weak matter-field coupling limit \cite{coleman_aspects_1985} suggests  that the phase diagram is shared by
  two phases characteristic of the Higgs mechanism \cite{anderson_gauge_63, englert_broken_64, higgs_broken_64, Guralnik_conservation_64}, a superfluid phase 1 with the quasi-condensation of bosons (Higgs phase) and a Mott-insulating phase 2 with strong interactions. 
However, 
non-perturbative calculations  show that the phenomenology in the phase 1 is the same as in the phase 2, and bosons are always tightly confined \cite{coleman_aspects_1985} (for a recent discussion 
see  \cite{komargodski_comments_2019, _david_a}).

One can 
certify the presence of a   phase transition in $d+1$ dimension for any $d > 2$ \cite{fradkin_phase_1979, Callaway_phase_1982}, but due to the  (boring) expectation of  a  single phase in the continuum,  the phase diagram of  AHM${}_2$ 
has never been computed on the lattice in 1+1D. This work  aims at filling this gap.

Our work is  strongly 
motivated by the current 
prospects of simulating lattice gauge theories using cold atomic setups \cite{dutta_toolbox_2017, Schweizer_2019, Gorg_2019, Mil_2020,Banuls2020, Aidelsburger2021}. Since
bosons are easier to cool down than fermions, 
 experiments along the lines proposed in \cite{kasamatsu2013,kuno2017,gonzalez-cuadra2017,zhang_quantum_2018,unmuth-yockey_universal_2018,park2019, Meurice_2021}
should  soon  explore the  phase diagram of  AHM${}_2$. 

\begin{figure}[t]
\includegraphics[width=\linewidth]{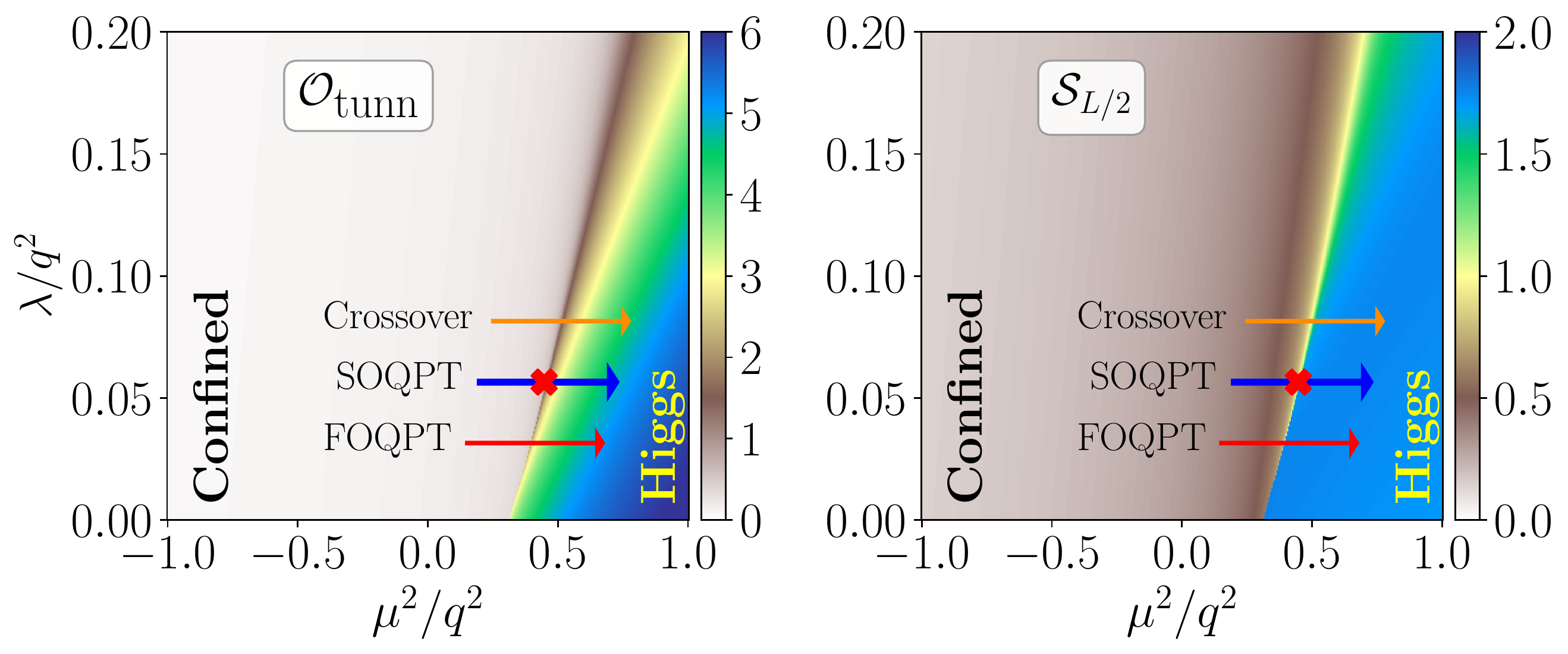}
\caption{Phase diagram of the AHM$_2$ \eqref{eq:Hamil} in the $(\mu^2/q^2, \lambda/q^2)$-plane for a system of size $L=60$.
At small couplings, the system occupies two qualitatively different regions, a confined and a Higgs region,  separated by a line of FOQPT as witnessed  by the average tunneling amplitude $\mathcal{O}_{\text{tunn}}$ (left panel) (effectively zero in the confined region, and finite in the Higgs region) and the entanglement entropy $\mathcal{S}_{L/2}$ measured at the center of the chain (right panel) (small in the confined region and large in the Higgs region). The line of FOQPT ends at a SOQPT, above which two regions are smoothly connected representing different aspects of a single phase.}
\label{fig:phase_diagram}
\end{figure}

Physicists have been working hard to measure the Higgs mode in experiments with cold atoms for a long-time, as reviewed in \cite{pekker_amplitude_2015}. In 2+1D, 
an explicit particle-hole symmetry  protects the decay of the Higgs mode into Goldstone modes allowing a proper measurement of its mass. These conditions are only met at the tip of the lobe of the Mott insulator to superfluid transition in Bose-Hubbard systems \cite{schori_excitations_2004,stoferle_transition_2004,endres_higgs_2012} which is, unfortunately, in 1+1D of the
Berezinskii-Kosterlitz-Thouless  type  (see e.g.,  \cite{kuhner_phases_1998, danshita_superfluidtomottinsulator_2011, Cazalilla_review_2011, Dutta_review_2015}) and is not particle-hole symmetric. This observation seems to  strengthen the picture emerging from the presence of a single phase in AHM${}_2$ and seems to suggest that a proper Higgs mode does not exist in 1+1D.

The results we present indicate a different picture, still characterized by a single phase, but with a reach landscape of transitions.

By performing matrix product states (MPS) \cite{schollwock_aop_2011, orus_aop_2014} simulations of the Hamiltonian version of  AHM${}_2$,
we  confirm the presence of a single phase for all the values of the mass of the bosons $\mu^2/q^2$ and their interaction strength $\lambda/q^2$ (in unit of the bosonic charge $q$)  in agreement with the field theoretical analysis.
However, unexpectedly,  for small $\lambda/q^2$ we find a line of first order quantum phase transitions (FOQPT) between the ``Higgs'' and the ``confined'' regions.
This line ends, for a finite value of $\lambda/q^2$, in a critical second order quantum phase transition (SOQPT), above which the two regions are continuously connected through a smooth crossover. Close to FOQPT line the two regions are well separated enabling identification of a ``Higgs'' mode and its analysis in the continuum limit (for a sufficiently small  $\lambda/q^2$). One can indeed take a different continuum limit to the standard one 
by approaching, from the Higgs region, the newly discovered SOQPT.

We 
precisely identify the  position and the nature of the new critical point.
By assuming Lorentz invariance at the critical point and then using the machinery of conformal field theories (CFT), we can understand the critical point as the direct sum of two non-interacting fields: a free fermionic field describing the Higgs mode and a free bosonic field, a collective mode of the Goldstone modes and the gauge field. 
Still the  a complete characterization of this critical point remains an outstanding challenge as some results do not fit the above picture.

\paragraph{The model.}
\label{sec:model}

Following \cite{chanda20}, we discretize  AHM$_2$ on a finite 1D lattice with   $L$ sites (with spacing $a$) (see \cite{supple} for details) arriving at the Hamiltonian (with open boundary conditions):
\begin{align}
\hat{H} = \sum_j \Big[ \hat{L}^2_j + 2 x \ \hat{\Pi}^{\dagger}_j \hat{\Pi}_j + (4 x - \frac{2 \mu^2}{q^2}) \hat{\phi}^{\dagger}_j\hat{\phi}_j  \nonumber\\ + \frac{\lambda}{q^2} (\hat{\phi}_j^{\dagger})^2 \hat{\phi}_j^2  - 2 x (\hat{\phi}^{\dagger}_{j+1} \hat{U}_j \hat{\phi}_j + h.c.)\Big],
\label{eq:Hamil}
\end{align}
 with  $x = 1/a^2 q^2$. The 
 matter fields $\{\hat{\phi}_j,  \hat{\phi}_j^{\dagger},
\hat{\Pi}_j,  \hat{\Pi}_j^{\dagger}\}$ operators act in Hilbert space at sites $j$, while the gauge-field $\{\hat{L}_j,  \hat{U}_j, \hat{U}^{\dagger}_j\}$ objects act in  Hilbert space defined on the bond linking sites $j$ and $j+1$.
The operators fulfill the standard commutation relations
$\ [\hat{\phi}_j, \hat{\Pi}_k] = [\hat{\phi}^{\dagger}_j, \hat{\Pi}^{\dagger}_k] = i \delta_{jk}$,
$[\hat{L}_j, \hat{U}_j] = -\hat{U}_j$ and $\ [\hat{L}_j, \hat{U}^{\dagger}_j] = \hat{U}^{\dagger}_j$.

The usual continuum limit 
is $x \rightarrow \infty$. Here we fix $x=2$ and characterize the phase diagram on the lattice.

We can define creation and annihilation operators for particles `$a$'  and anti-particles `$b$'  as $\hat{a}_j$ and $\hat{b}_j$ fulfilling
$[\hat{a}_j, \hat{a}^{\dagger}_k] = [\hat{b}_j, \hat{b}^{\dagger}_k] = \delta_{jk}$ \footnote{The operators are defined as
\begin{eqnarray*}
\hat{\phi}_j = \frac{1}{\sqrt{2}} \left(\hat{a}_j + \hat{b}_j^{\dagger}\right), \ \hat{\Pi}_j = \frac{i}{\sqrt{2}} \left(\hat{a}^{\dagger}_j - \hat{b}_j\right), \\
\hat{\phi}^{\dagger}_j = \frac{1}{\sqrt{2}} \left(\hat{a}^{\dagger}_j + \hat{b}_j\right), \ \hat{\Pi}^{\dagger}_j = \frac{i}{\sqrt{2}} \left(\hat{b}^{\dagger}_j - \hat{a}_j\right),
\end{eqnarray*} as discussed in e.g., \cite{chanda20}.}.
 We 
 use the density matrix renormalization group (DMRG) algorithm \cite{schollwock_aop_2011,orus_aop_2014,white_prl_1992,white_prb_1993,white_prb_2005,schollwock_rmp_2005}
 to find the ground state of the  Hamiltonian \eqref{eq:Hamil}.  Specifically, we employ a strictly single-site variant of DMRG with subspace expansion \cite{Hubig_prb_2015}.
For numerics we  limit the occupations of bosonic modes to at most $n_0^a = n_0^b = 10$
\footnote{For details about the DMRG simulations, see \cite{supple}.}.

In absence of external charges, the local $\mathbb{U}(1)$ symmetry implies the  Gauss law  $\hat{G}_j =0,\ \forall j$, where the generators are \cite{chanda20}
$
\hat{G}_j = \hat{L}_j - \hat{L}_{j-1} -{\hat{Q}_j},
$
and $\hat{Q}_j = \hat{a}_j^{\dagger} \hat{a}_j - \hat{b}_j^{\dagger} \hat{b}_j$ encodes the density of dynamical charges.
Using the Gauss law, we can integrate-out the gauge-fields in a chain  with open-boundary conditions in favor of a long-range potential for the matter fields \cite{schwinger_pr_1951}.

\paragraph{Confined and Higgs, two shades of the same phase.} 
The  long-range interactions among bosons destroy the  phases of the standard 1+1D Bose Hubbard model -- see
 the phase diagram 
 in  Fig.~\ref{fig:phase_diagram}.
  For $\lambda/q^2 \geq 0$, the system is in the confined region as far as $\mu^2/q^2 \leq 0$. In this region the model has a finite mass gap, and the elementary excitations are mesons,  bound pairs of  particle-antiparticle. The gauge bosons are in the lowest eigenstate of  $\hat{L}_j$, so that the variance $\sigma^2(\hat{L}_j) = \braket{\hat{L}^2_j} - \braket{\hat{L}_i}^2 \approx 0$.

For $\mu^2/q^2 \gg 0$,  the system enters the \textit{gapped} Higgs region where the variance $\sigma^2(\hat{L}_j)$ becomes large. 
The effective gauge-field mediated tunneling amplitude $\mathcal{O}_{\text{tunn}} = \frac{1}{2L} \sum_j \braket{\hat{\phi}^{\dagger}_{j+1} \hat{U}_j \hat{\phi}_j + h.c.}$ increases, so that it can distinguish the Higgs region from the confined one.

We can also characterize the two regions by  considering the behavior of the entanglement entropy of a block made of   $l$ constituents starting from the boundary, defined as
\begin{equation}
\mathcal{S}_l = -\text{Tr} \left[ \rho_l \ln \rho_l \right],
\end{equation}
where $\rho_l = \text{Tr}_{l+1, l+2, ...,L} \ket{\psi}\bra{\psi}$ is the reduced density matrix. In the Higgs region, the entanglement entropy is systematically larger than in the confined region. However, since both phases are gapped, the entropy follows area-law scaling with respect to the bipartition size (see \cite{supple}).


For  sufficiently small $\lambda/q^2$,  the two regions are separated by a  FOQPT line characterized by discontinuous jumps both in the 
tunneling amplitude $\mathcal{O}_{\text{tunn}}$ and in 
$\mathcal{S}_{L/2}$ measured across the central bond \cite{supple}. This line terminates at a critical SOQPT at a finite value of $\lambda_c/q^2 $ and $\mu_c^2/q^2$, identified by the red cross in Fig.~\ref{fig:phase_diagram}. We discuss  precise location and characterization of this critical point below.
Above the SOQPT,  the two regions are smoothly connected as revealed by smooth changes in all the physical quantities while moving from one region to the another.

\paragraph{Nature of the critical point.} We can precisely locate and characterize
 the critical point assuming its Lorentz invariance, 
that implies an applicability of a CFT at low energies.
In a CFT, the finite-size scaling of the entanglement entropy of a block of  first $l$ consecutive sites in a chain with open boundary conditions and length $L$ is
\begin{equation}
\mathcal{S}(l,L) = \frac{c}{6} W + b',
\label{eq:cardy_calabrese}
\end{equation}
where $c$ is the central charge of the corresponding CFT, $b'$ is a non-universal constant and the chord length $W$ is a function of both $L$ and $l$:
$
W(l,L) =\ln \left[   \frac{2 L}{\pi} \sin(\pi l /L)\right]$~\cite{callan_geometric_1994, vidal_2003, calabrese_entanglement_2004}.

\begin{figure}[h]
\includegraphics[width=\linewidth]{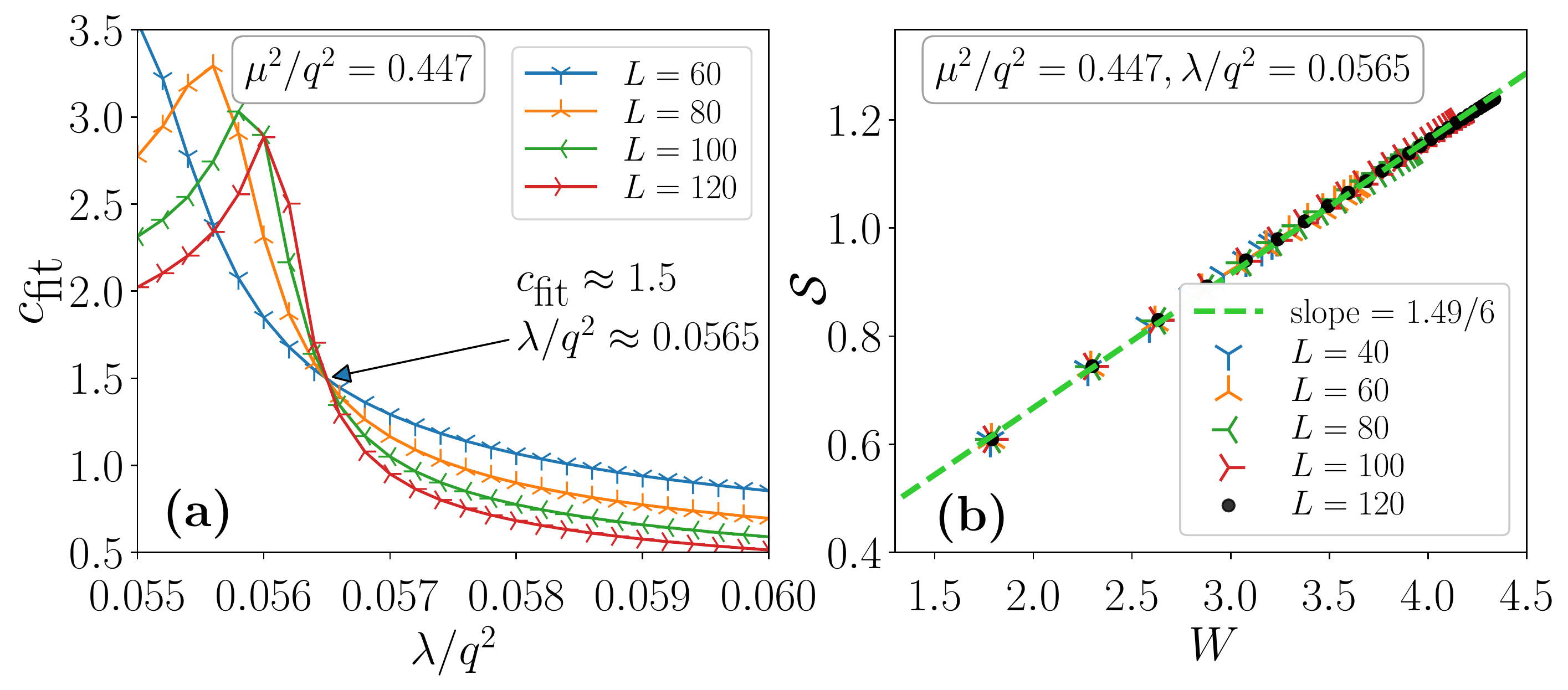}
\caption{Entropy scaling of the AHM${}_2$. \textbf{(a)} The fitted central charge $c_{\text{fit}}$ according to \eqref{eq:cardy_calabrese} for fixed $\mu^2/q^2 = 0.447$ and different system sizes. Curves for different system-sizes crosses each other at $\lambda/q^2 \approx 0.0565$ and $c_{\text{fit}} \approx 3/2$.
\textbf{(b)} The scaling of the entanglement entropy at the critical point for different system sizes yields  the central charge of the critical theory as $c = 1.49(1)$.\label{fig:central_charge_u1}}
\end{figure}

 We  pin-point the SOQPT by adapting the idea of  the phenomenological renormalization group \cite{nightingale_scaling_1975} to the scaling of the entropy in  \eqref{eq:cardy_calabrese} as  explained in  \cite{koffel_entanglement_2012,buyskikh_spin_2019}. At the critical point,  the $c$ value should be independent of the system's size. For each  $L$, we obtain $c_{\text{fit}}$ by fitting  \eqref{eq:cardy_calabrese} to our numerical data for $\mathcal{S}(l,L)$. The extracted values in the $(\mu^2/q^2, \lambda/q^2)$-plane depend on $L$ and become independent of the system size only at the critical point. Our data suggest that the $L$ dependent central charges collapse to a single value at $(\mu^2/q^2 = 0.447(1), \lambda/q^2 = 0.0565(1))$  (see Fig.~\ref{fig:central_charge_u1}).
 The central charge at the critical point $(\mu_c^2/q^2 = 0.447, \lambda_c/q^2 = 0.0565)$ is found to be  $c = 1.49(1)$
 \footnote{Notice that our analysis is based on the full scaling form of the entropy in \eqref{eq:cardy_calabrese} that holds for conformally invariant systems only. The logarithmic divergence of the half chain entropy on the other hand can be observed also for systems that only  posses scale invariance \cite{refael_entanglement_2004, latorre_entanglement_2005,koffel_entanglement_2012}.}.

The value of the central charge mentioned above 
suggests that we are not dealing with a minimal model. However, we want to argue here that we are in the presence of the direct sum of two different minimal models, each contributing to a piece of the total central charge,  a $c_f=1/2$ for a free Majorana fermion and a $c_b=1$ for free boson.
This scenario is strongly motivated by the standard Higgs mechanism. The complex Higgs field separates into its amplitude and its phase. The amplitude mode is effectively described by a real $\lambda \phi^4$ theory that undergoes the standard Ising phase transition (the $c=1/2$ part). The phase, on the other hand, provides the longitudinal degree of freedom to the photon field. The latter becomes massless at the transition and provides the $c=1$ free bosonic part. The value of $c=1.5$ furthermore  suggests, based on the $c$ theorem, that the two parts should be non-interacting \cite{zomolodchikov_irreversibility_1986}.

\begin{figure}[b]
\centering
\includegraphics[width=\linewidth]{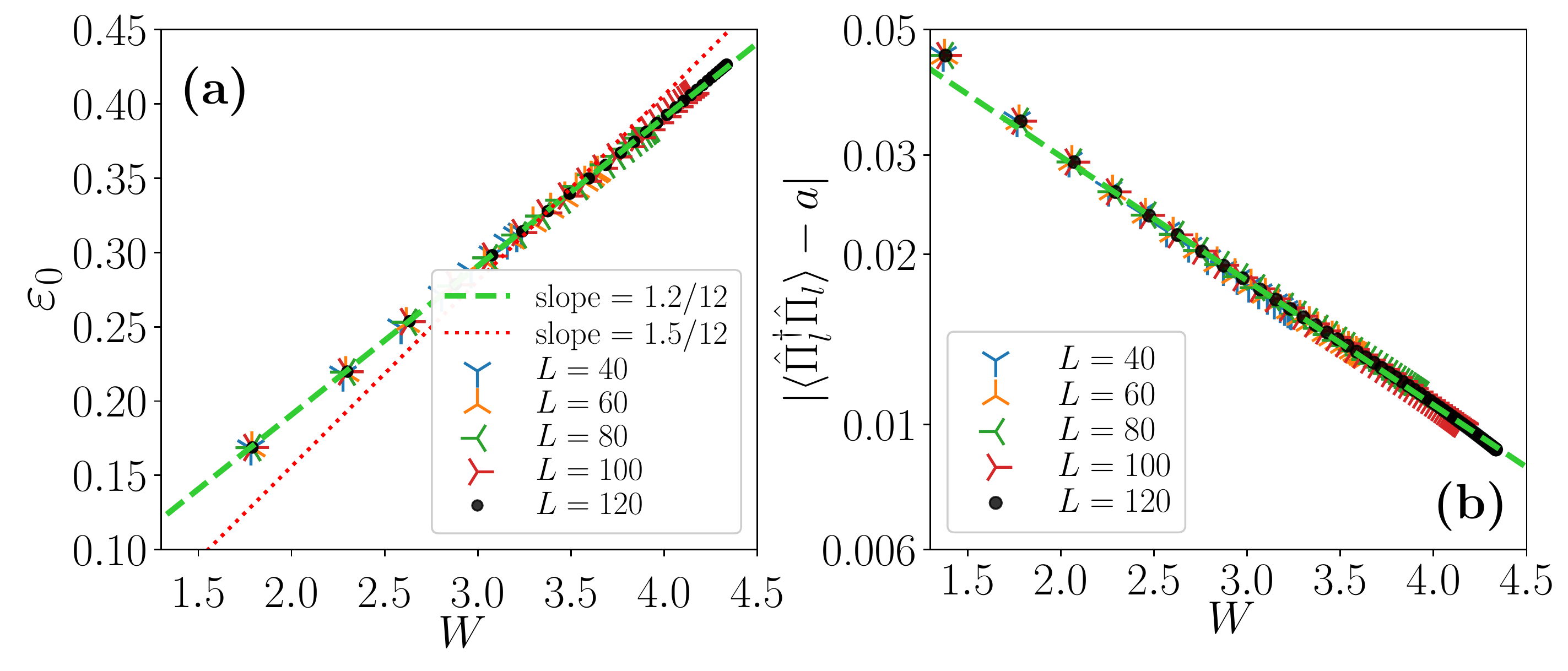}
\caption{
\textbf{(a)} The scaling of the entanglement ground state $\varepsilon_0$ matches perfectly the functional form suggested by the CFT analysis, reported in the  text. However, the numerical value for the central charge deviates by around the $20\%$ from the value we extract from the scaling of the entanglement entropy as the best fit suggests $c_{\text{eff}}=1.20(1)$.
The red-dotted line depicts the fit assuming $c=1.5$.
\textbf{(b)} The scaling of $\braket{\hat{\Pi}^{\dagger}_l\hat{\Pi}_l}$ according to the CFT prediction (Eq. \eqref{eq:pipi}). 
It couples both to the identity operator and one primary with scaling dimension $\Delta$ that comes out to be  $\Delta =0.51(2)$ from the fit, with $a$ being $0.5474(2)$}.
\label{fig:L2_PidagPi}
\end{figure}

In order to confirm this scenario we compute the entanglement spectrum that is also known to encode the central charge of the theory \cite{cardy_entanglement_2016a}. The entanglement spectrum, denoted by $\varepsilon_s$, is the spectrum of the entanglement Hamiltonian $H_l =-\log(\rho_l)$.
By assuming a factorized ground state, we should observe that the smallest eigenvalue of $H_l$, $\varepsilon_0$ diverge logarithmically. In particular we should see that \cite{cardy_entanglement_2016a} 
\begin{equation}
\varepsilon_0 = \left(\varepsilon_0^{Ising}+\varepsilon_0^{boson} \right) \propto \frac{(c_f+c_b)}{12} W + O(1/W).
\label{eq:entSpec0}
\end{equation}
By fitting our numerical data to \eqref{eq:entSpec0}, we observe a perfect collapse on the functional form predicted by CFT, but the numerical result for of $c_{\text{eff}} = 1.20(1)$ is not compatible with  $1.5$ (see Fig.~\ref{fig:L2_PidagPi}\textbf{(a)}). This disagreement between the scaling of the entanglement entropy and that for the entanglement ground state already suggests that the critical point is unusual and exotic in nature.

We thus turn to analyze the operator content of the model by studying the correlation functions of local operators. We should be able to identify a set of primary operators, by studying the large distance  two point correlations function that should decay algebraically as $\phi(0)\phi(r) \sim 1/r^{{\Delta_{\phi}}}$.
The presence of gauge symmetry, however, strongly reduces the set of operators we can consider.  Most of the candidates that should couple to primary operators  are either trivial (due to the low-dimensionality of the system) or vanishing since they are not gauge invariant. The only non-vanishing operators are indeed Wilson lines terminating on a boson-antiboson pair, and electric field correlations. We also have access to local operators such as $\hat{\phi}^{\dagger}\hat{\phi}$ and $\hat{\Pi}^{\dagger}\hat{\Pi}$ that couple both to the real part and the phase of the field.
By assuming we are dealing with a CFT, we can use  the  conformal map that  maps  the profile of local operators to two points correlation functions on the full plane (see e.g. \cite{cardy_introduction_1988}).

At first we analyze the behavior of $\braket{\hat{L}^2_l}$  as function of the chord coordinate $W$. The numerical results show that $\braket{\hat{L}^2_l}$
diverges linearly as a function of $W$,  unveiling that $\hat{L}^2$ behaves as a free-bosonic field. Furthermore, 
the slope of such linear scaling is found to be $1.20(4)/12$, matching that of the entanglement ground state energy.
Turning to analyzing the profile of $\braket{\hat{\Pi}^{\dagger}_l\hat{\Pi}_l}$ as a function of $W$ we find that
\begin{equation}
\braket{\hat{\Pi}^{\dagger}_l\hat{\Pi}_l} \simeq a+b(\exp(W))^{-\Delta},
\label{eq:pipi}
\end{equation}
where $a$ and $b$ encode the overlap of the above expectation value with the identity operator and one of the primaries. The numerical data (Fig.~\ref{fig:L2_PidagPi}\textbf{(b)}) suggest that $\Delta \simeq 0.5$, the conjugate operator to  the one that would match to the derivative of the electric field.

Unfortunately, we do not  find any operator that couples to the primary of the Ising part of the CFT. Summarizing,  our data seem  to confirm that we have one part of the system that behaves as  free boson and suggest that  $\partial_x \hat{L}^2$ should have a large overlap with the primary operator (the derivative of the free bosonic field), while $\hat{\Pi}^{\dagger}\hat{\Pi}$ should have a strong overlap with the conjugate primary operator.

\begin{figure}[t]
\includegraphics[width=\linewidth]{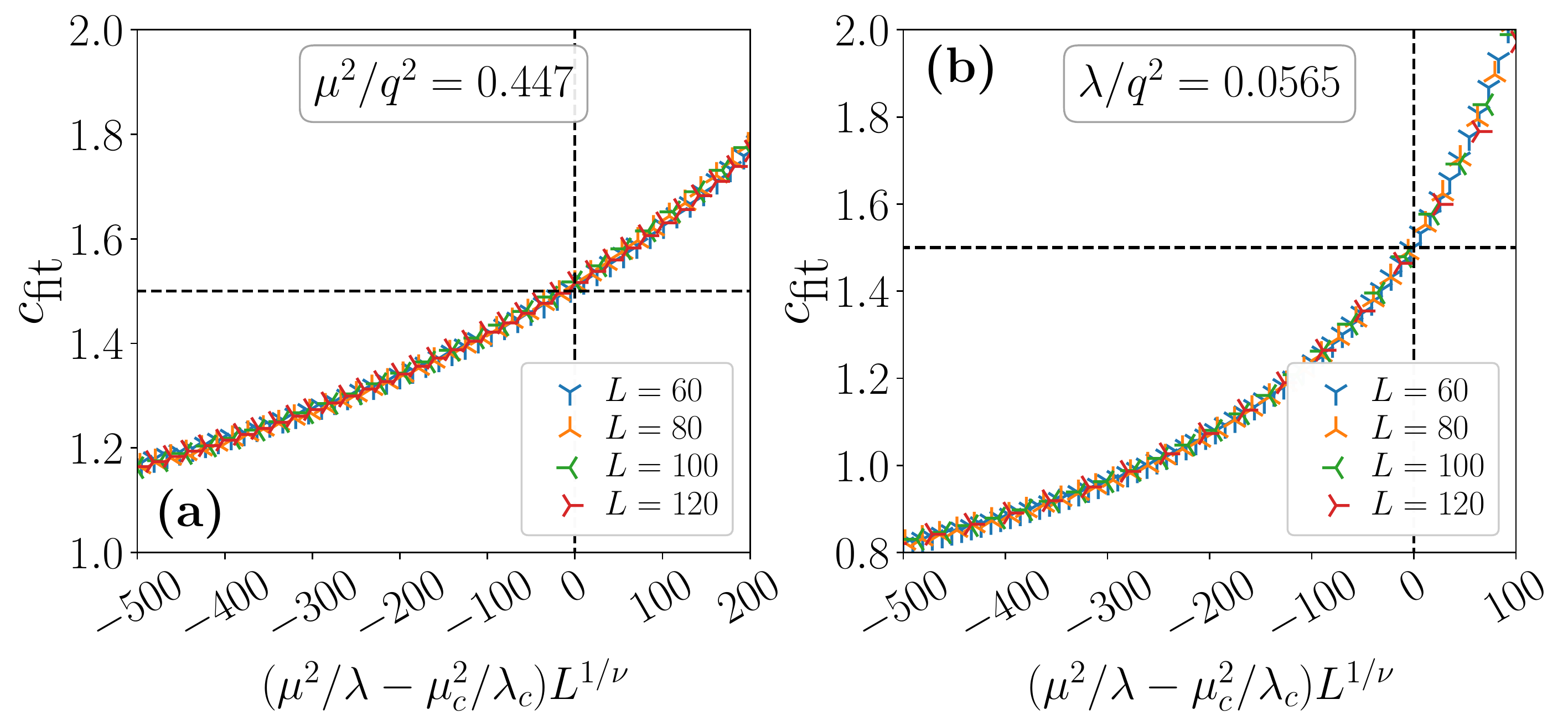}
\caption{
The collapse of $c_{\text{fit}}$ according to the scaling hypothesis \eqref{eq:collapse} in the neighborhood of the critical point $\mu_c^2/\lambda_c = 0.447/0.0565$
for fixed $\mu^2/q^2 = 0.447$ (left) and for fixed $\lambda/q^2 = 0.0565$ (right). 
Here, we vary \textbf{(a)} $\lambda/q^2$ in the range $[0.056, 0.058]$ and \textbf{(b)} $\mu^2/q^2$ in the range $[0.435, 0.45]$ for the data collapses.
In both the cases, the critical exponent is found to be $\nu = 0.5 \pm 0.02$ from data collapses.
\label{fig:critical_expo}
}
\end{figure}

Now moving away from the critical point, we can use the standard scaling hypothesis to extract the exponent $\nu$ from the collapse of the fitted central charge as
\begin{eqnarray}
c_{\text{fit}}(L) = f\left((\mu^2/\lambda - \mu_c^2/\lambda_c)L^{1/\nu}\right),
\label{eq:collapse}
\end{eqnarray}
where $f(.)$ is a continuous function and $\nu$ is the corresponding critical exponent.
Performing the data collapse according to \eqref{eq:collapse} in the neighborhood of the critical point $\mu_c^2/\lambda_c$ (see Fig.~\ref{fig:critical_expo})
we find the critical exponent to be $\nu = 1/2 \pm 0.02$ that matches the value observed in the transition  from polarized to critical phase in the XX model in a magnetic field  \cite{latorre_ground_2004, campostrini_quantum_2010, dutta_book_2015}.
The same transition can be understood in terms of free bosons that pass from their Fock vacuum to the superfluid regime as the chemical potential exceeds the width of the first band. In our case, the strange thing is that there is no superfluid regime, but just a single critical point where the gauge-boson condense, while away from the critical point our system passes from vacuum to a Mott insulator phase.
Now using the standard scaling hypothesis, once we have figured out that $\nu=1/2$ we can deduce that $\braket{\hat{\Pi}^{\dagger}\hat{\Pi}} \simeq (\mu^2/\lambda - \mu_c^2/\lambda_c)$, meaning that $\beta=1$.

It is worth pointing out that $\nu=1/2$, seems inconsistent with a CFT, where we would expect, that $d-1/\nu=\Delta$ with $\Delta$ the thermal critical exponent and $d$ being 2 for the 1+1D quantum system, while we find $d-1/\nu=0$. However, all our results so far have been obtained by assuming a full conformal invariance in mapping the correlation function of our finite system to the ones of an infinite plane by means of a conformal transformations.

The appearance of $\nu=1/2$, together with the failure to identify a local operator that couples to the primary field of the Ising part of the CFT, contrasts with the factorization on the critical point. However, by repeating a similar analysis in a $\mathbb{Z}_3$ gauge theory coupled to bosonic matter we find a $c \simeq 0.8+0.5$  \cite{chanda2021}. As a result we still believe that the factorization hypothesis is correct, but it requires a further analysis to be appropriately confirmed. In particular, it would be interesting to analyze the system under periodic boundary condition, which, however, is impractical at current computational capabilities using MPS ansatz but may become a possibility using next-generation tensor network algorithms.

\paragraph{Discussion and Conclusions.}
\label{sec:conclu}
We have analyzed the phase diagram of AHM$_2$ on a discrete lattice in 1+1D. We have found two distinct regions, the confined and the Higgs regions that are separated by line of FOQPT that terminates at a SOQPT. Beyond the SOQPT the two regions are connected by a smooth crossover.
The presence of a SOQPT allows one to construct an unorthodox continuum limit of the theory that  should be described  by  free fermions and  free bosons that do not interact.

This would result in a CFT with central charge $c=3/2$, compatible with our numerical result and would have a compelling  interpretation in terms of the standard Higgs mechanism --  the real part of the complex field undergoes an Ising transition (the $c=1/2$ part), while the phase of it provides the transverse degree of freedom to the photon that becomes dynamical and massless (the $c=1$ part). 

However, further numerical analyses unveil surprizing pieces of the puzzle
that do not fit our interpretation.We did not  find a local operator that couples to the $c=1/2$ part of the CFT. The scaling of the entanglement ground state should follow a similar law to the one of the entanglement entropy.  The numerical value of the central charge that we extract from it is $c=1.20(1)$. We also obtain  $\nu=1/2$ analyzing the collapse of the data for the entanglement entropy close to the critical point, in contrast to the expected $\nu=1$.

Are we actually observing a Lorentz invariant critical point where the Higgs and photon mode factorize?  We believe this is the case 
as also supported by the presence of a linear dispersion relation witnessed by the non-zero ``sound velocity'' extracted from a finite-size scaling analysis of the ground state energy \cite{supple}.
Still our study leaves some questions unanswered. We strongly believe that our paper will open the debate, and that together with the broader scientific community we will soon have a final picture of the mechanism behind this newly observed critical point.


\acknowledgments
We are grateful to Marcello Dalmonte for the valuable suggestion regarding the analysis of the sound velocity.
T.C. thanks Subhroneel Chakrabarti for useful discussions.
L.T. would like to acknowledge the discussions with F. Gliozzi, B. Fiol, and  E. Vicari on the topics presented.
The numerical computations have been possible thanks to   PL-Grid Infrastructure.
The works of T.C. and J.Z have been realized within the QuantERA grant QTFLAG, financed by National Science Centre (Poland)  via grant 2017/25/Z/ST2/03029.
 M.L. acknowledges support from ERC AdG NOQIA, State Research Agency AEI (``Severo Ochoa'' Center of Excellence CEX2019-000910-S, Plan National FIDEUA PID2019-106901GB-I00/10.13039 / 501100011033, FPI, QUANTERA MAQS PCI2019-111828-2 / 10.13039/501100011033), Fundació Privada Cellex, Fundació Mir-Puig, Generalitat de Catalunya (AGAUR Grant No. 2017 SGR 1341, CERCA program, QuantumCAT U16-011424, co-funded by ERDF Operational Program of Catalonia 2014-2020), EU Horizon 2020 FET-OPEN OPTOLogic (Grant No. 899794), and the National Science Centre, Poland (Symfonia Grant No. 2016/20/W/ST4/00314), Marie Sk{\l}odowska-Curie grant STRETCH No. 101029393, “La Caixa” Junior Leaders fellowships (ID100010434),  and EU Horizon 2020 under Marie Sk{\l}odowska-Curie grant agreement No 847648 (LCF/BQ/PI19/11690013, LCF/BQ/PI20/11760031,  LCF/BQ/PR20/11770012).
L.T. acknowledges support from the Ram\'on y Cajal program RYC-2016-20594, the ``Plan Nacional Generaci\'on de Conocimiento'' PGC2018-095862-B-C22, the State Agency for Research of the Spanish Ministry of Science and Innovation through the ``Unit of Excellence María de Maeztu 2020-2023'' award to the Institute of Cosmos Sciences (CEX2019-000918-M), and the European Union Regional Development Fund within the ERDF Operational Program of Catalunya, Spain (project QUASICAT/QuantumCat, ref. 001-P-001644).

\bibliographystyle{apsrev4-1}

\bibliography{higlet4.bbl}

\clearpage
\onecolumngrid
\begin{center}
  \textbf{\large Supplementary Material: \\
Phase diagram of 1+1D Abelian-Higgs model and its critical point}\\[1cm]
\end{center}

\twocolumngrid

\subsection{From the continuum to the lattice}

The Lagrangian density of the Abelian-Higgs model is  given by \cite{peskin_book, _david_a}
\begin{eqnarray}
\mathcal{L}&=& -\left[ D_{\mu} \phi \right]^* D^{\mu} \phi  - \frac{1}{4}F_{\mu\nu}F^{\mu\nu} - V(\phi),   
\label{eq:lagrangian} 
\end{eqnarray}
with 
\begin{eqnarray}
V(\phi) =  - \mu^2 |\phi|^2 + \frac{\lambda}{2} |\phi|^4. 
\end{eqnarray}
Here $\phi$ is the complex scalar field, $D_{\mu} = (\partial_{\mu} + i q A_{\mu})$ is the covariant derivative with $q$ and $A_{\mu}$ being the gauge coupling and the electromagnetic vector potential respectively,  and $F_{\mu\nu} = \partial_{\mu} A_{\nu} - \partial_{\nu} A_{\mu}$ is the electromagnetic field tensor. 
Here, we use the metric convection $(-1,1,1,1)$ or $(-1,1)$ (in 1+1 dimension).  

For $\mu^2 < 0$, the system describes the bosonic Schwinger model (BSM) \cite{chanda20} with added $|\phi|^4$-interaction, while $\mu^2 > 0$ describes Abelian-Higgs model (AHM) where the potential attains minimum ($V_{\min} = - \frac{\mu^4}{2 \lambda}$) at non-zero value of the field-strength $|\phi_0| = \sqrt{\mu^2/\lambda}$, leading to the spontaneous symmetry breaking at semiclassical level.

After fixing the temporal gauge $A_t(x, t) = 0$, we get the quantum 1+1D Hamiltonian (AHM$_2$) in the continuum as 
\small
\begin{eqnarray}
\hat{H} &=& \int dx \bigg[ \hat{\Pi}^{\dagger}(x) \hat{\Pi}(x) -  \mu^2 \hat{\phi}^{\dagger}(x) \hat{\phi}(x) + \frac{\lambda}{2} (\hat{\phi}^{\dagger}(x))^2 (\hat{\phi}(x))^2 + \nonumber \\
&+& \frac{1}{2} \hat{E}_x^2(x) + \left(\partial_x - i q \hat{A}_x(x)\right) \hat{\phi}^{\dagger}(x) \left(\partial_x + i q \hat{A}_x(x)\right) \hat{\phi}(x) 
\bigg], \nonumber \\
\end{eqnarray}
\normalsize
where $\hat{E}_x(x)$, $\hat{\Pi}(x)$, and $\hat{\Pi}^{\dagger}(x)$ are the canonical conjugate operators corresponding to $\hat{A}_x(x)$, $\hat{\phi}(x)$, and $\hat{\phi}^{\dagger}(x)$ respectively,
satisfying the canonical commutation relations:
\begin{align}
[\hat{A}_x(x_1), \hat{E}_x(x_2)]  &=  [\hat{\phi}(x_1), \hat{\Pi}(x_2)]  \nonumber \\
&=  [\hat{\phi}^{\dagger}(x_1),  \hat{\Pi}^{\dagger}(x_2)] = i \delta(x_1-x_2). 
\end{align}

\noindent \textbf{Note:} The quantization of the $|\phi|^4$ term may be done in various manners. Here we do not enforce any normal ordering but rather take $|\phi|^4 \xrightarrow[]{\text{quantization}} (\phi^\dagger)^2 \phi^2$.

Following \cite{chanda20}, we can straightforwardly discretize the above Hamiltonian on a 1D spatial lattice with the lattice spacing $a$, and we ultimately arrive at the Hamiltonian \textcolor{blue}{(1)} in the main text.

\begin{figure}
\centering
\includegraphics[width=\linewidth]{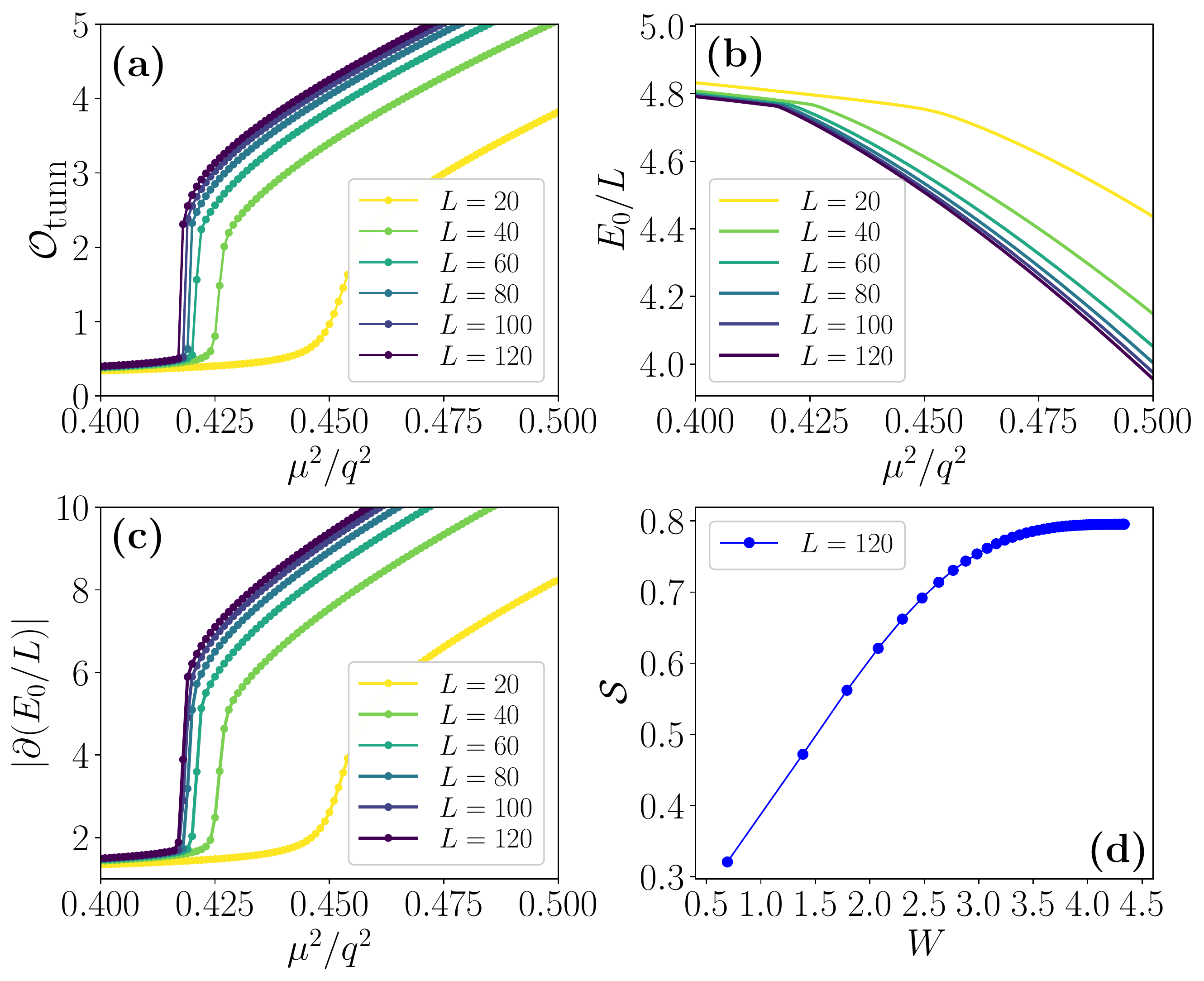}
\caption{The study of the first order quantum phase transition at fixed $\lambda/q^2=0.045$ and varying $\mu^2/q^2$ for different system-sizes $L \in [20, 120]$. \textbf{(a)}-\textbf{(c)} respectively show the variations of $\mathcal{O}_{\text{tunn}}$, the energy density and its derivative with respect to the system parameter $\mu^2/q^2$. \textbf{(d)} The entanglement scaling with respect to the cord length $W =  \ln \left(\frac{2L}{\pi} \sin(\pi l/L) \right)$ for $L=120$ at the first order transition that unveils the expected area-law scaling of the entropy.
}
\label{fig:first_order}
\end{figure}

\subsection{The first order quantum phase transition}

As reported in the main text, below the critical interaction strength $\lambda_c/q^2 = 0.0565$ the confined and the Higgs regions are separated by a first order quantum phase transition (FOQPT) line. To clearly demonstrate the existence and features of the FOQPT, in Figs.~\ref{fig:first_order}  we present different observables across the transition as a function of $\mu^2/q^2$ for fixed $\lambda/q^2 < \lambda_c/q^2$. Specifically, we depict the behaviors of $\mathcal{O}_{\text{tunn}} = \frac{1}{2L} \sum_j \braket{\hat{\phi}^{\dagger}_{j+1} \hat{U}_j \hat{\phi}_j + h.c.}$, the energy density and its derivative with respect to the system parameter $\mu^2/q^2$ respectively in Figs.~\ref{fig:first_order}\textbf{(a)}-\textbf{(c)} for different system-sizes $L \in [20, 120]$.
The average tunneling amplitude $\mathcal{O}_{\text{tunn}}$ clearly shows discontinuous jumps as we vary  $\mu^2/q^2$ across the phase transition for sufficiently large system-sizes. 
On other hand, the ground state energy shows non-analytic kinks in its profile at the transition point, such that its derivative (with respect to the parameter $\mu^2/q^2$) manifests a discontinuity. 
To observe the discontinuous jumps or kinks in the profile of $\mathcal{O}_{\text{tunn}}$ or of the energy, we need sufficiently large system-size as  FOQPTs are associated with large (but finite) correlation lengths, as a result smaller system sizes cannot properly resolve the FOQPT.
In Fig.~\ref{fig:first_order}\textbf{(d)}, we also show the scaling of the entanglement entropy at the FOQPT that features the expected area-law behavior for bipartitions larger than the correlation length.

\subsection{Entanglement scaling in the gapped phases}

\begin{figure}
\centering
\includegraphics[width=\linewidth]{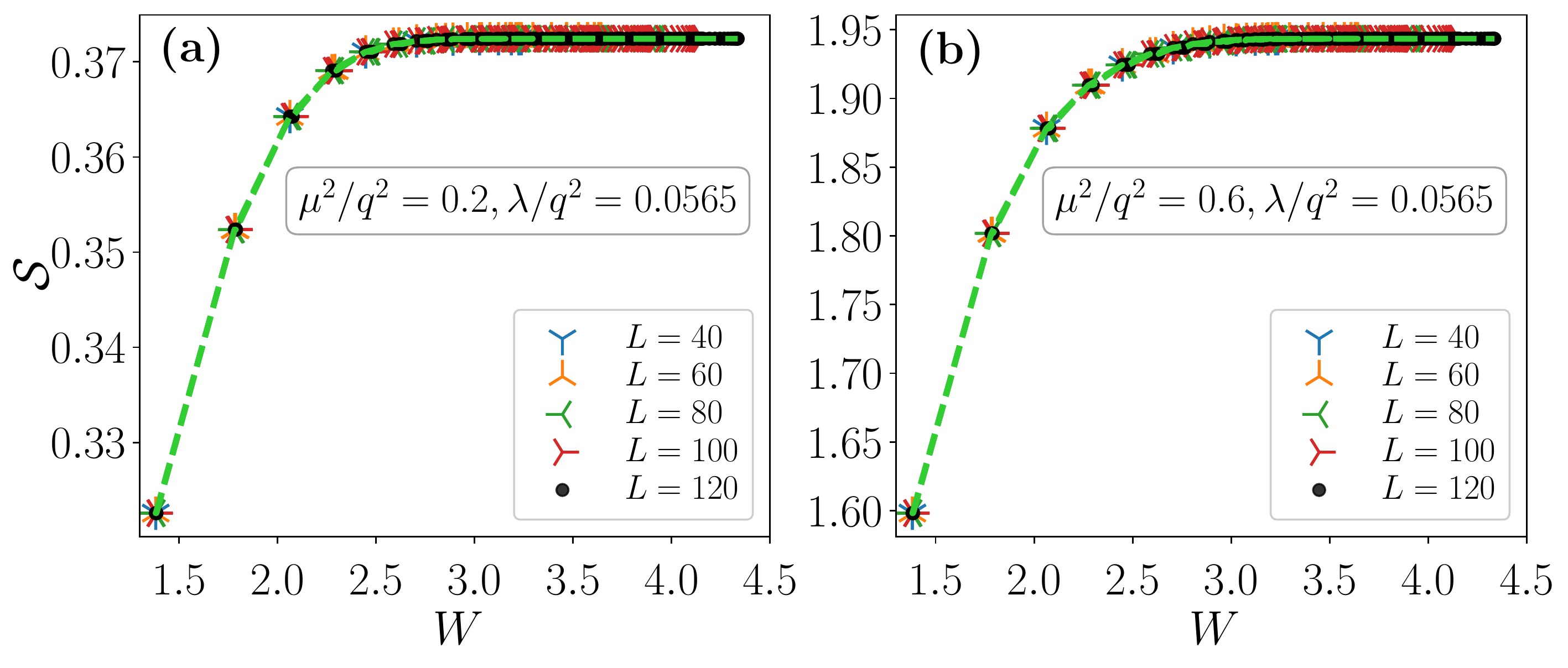}
\caption{The profile of the entanglement entropy with respect to the cord length $W$ at \textbf{(a)} the confined region and \textbf{(b)} the Higgs region.
\label{fig:entanglement_other_phases}
}
\end{figure}

As mentioned in the main text,
apart from the critical point, the Hamiltonian in every point of the phase diagram is gapped, therefore the entanglement entropy follows asymptotically, for large enough blocks,  the  standard area-law scaling. To show this, we plot the the entanglement entropy with respect to the cord length $W$ in the confined and the Higgs regions in Fig.~\ref{fig:entanglement_other_phases}. In both cases, the entanglement entropy becomes flat with increasing values of $W$, showing that for large enough blocks it saturates to a value that does not depend on the block size (area-law).

\begin{figure}
\includegraphics[width=\linewidth]{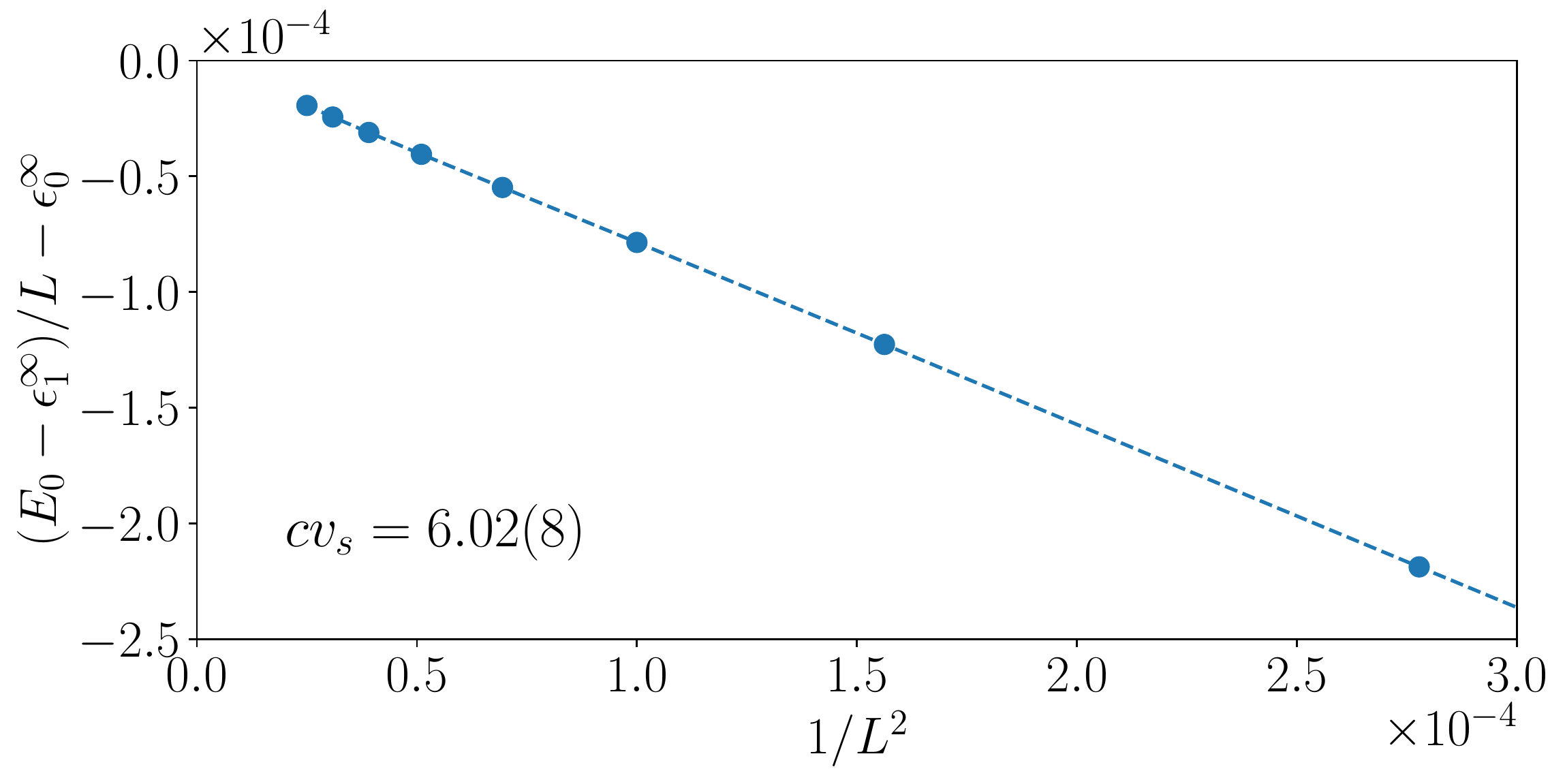}
\caption{
Finite-size scaling of the ground state energy $E_0$ at the critical point according to Eq.~\eqref{eq:gs_scaling} that gives the sound velocity $v_s  \simeq 4$.
\label{fig:gs_scaling}
}
\end{figure}

\subsection{Evidence for Lorentz invariance of the critical point}

In the main text, we have analyzed  the critical point $(\mu_c^2/q^2 = 0.447, \lambda_c/q^2 = 0.0565)$ by means of predictions from conformal field theory (CFT),  assuming Lorentz invariance. In order to verify this assumption, we calculate the ``sound velocity'' at the critical point from the scaling of ground state energy. For a Lorentz invariant system with open boundary condition (OBC) having a linear dispersion at low energies, the ground state energy $E_0$  scales with the system size $L$ as \cite{Blote_1986, Affleck_1986, Affleck_1989}
\begin{equation}
E_0(L) = \epsilon_0^{\infty} L + \epsilon_1^{\infty} - \frac{\pi c v_s}{24 L},
\label{eq:gs_scaling}
\end{equation}
where $\epsilon_0^{\infty}$ is the ground state energy density in the bulk and $\epsilon_1^{\infty}$  is the surface free energy in the thermodynamic limit, $c$ is the central charge of the corresponding CFT, and $v_s$ is the sound velocity. For a Lorentz invariant system $v_s$ must be non-zero, while it vanishes for Lorentz non-invariant critical points with a quadratic dispersion.

To estimate $v_s$, we perform the finite-size scaling of the ground state energy according to Eq.~\eqref{eq:gs_scaling} at the critical point (for similar analysis, see \cite{Hallberg_1996, xavier_2010, Dalmonte_2012, Chepiga_2017}). Such a finite-size scaling (Fig.~\ref{fig:gs_scaling}) yields $c v_s = 6.02 \pm 0.08$ ($v_s \simeq 4$ for $c = 1.5$), confirming the Lorentz invariance of the critical point analyzed in the main text.

\subsection{Details about numerical simulations}

\begin{figure}
\centering
\includegraphics[width=\linewidth]{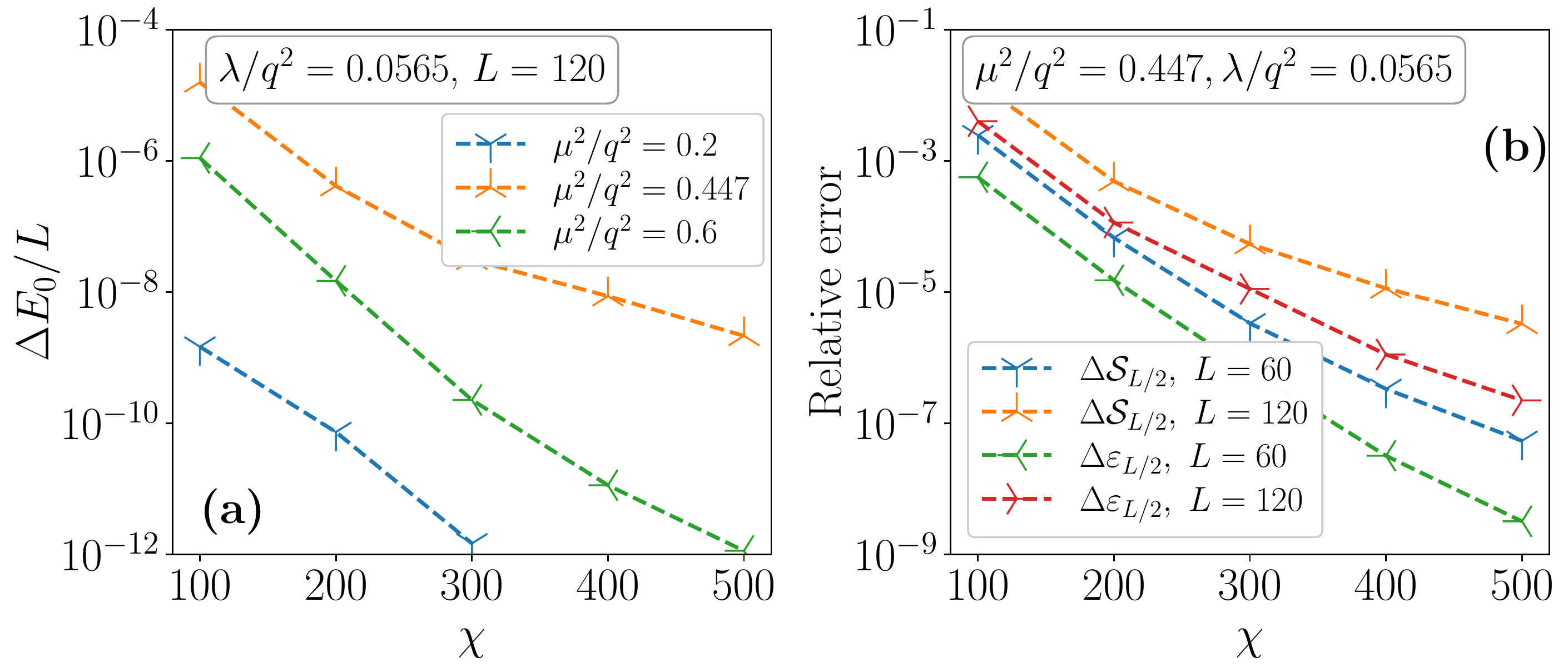}
\caption{The convergence of \textbf{(a)} the energy density ($\Delta E_0/L$) and \textbf{(b)} the half-chain entanglement entropy ($\Delta \mathcal{S}_{L/2}$) and the half-chain entanglement ground state energy ($\Delta \varepsilon_{L/2}$)  with respect to the maximum bond dimension used in the simulation, respectively $\chi \in \{100, 200, 300, 400, 500, 600\}$. Here we plot relative errors in the quantities as $\Delta \mathcal{O} = |\mathcal{O}_{\chi} - \mathcal{O}_{\chi+100}|$. In \textbf{(a)}  we consider three points in the phase diagram, namely (1) a point in the confined region ($\mu^2/q^2 = 0.2$), (2)  the critical point ($\mu^2/q^2 = 0.447$), and (3) a point in the Higgs region ($\mu^2/q^2 = 0.6$) for a system of size $L=120$. In \textbf{(b)} we show the convergence of the entropic quantities as a function of the bond dimension at the critical point.
\label{fig:chi_convergence}
}
\end{figure}

The results reported in the article have been obtained using  the density matrix renormalization group (DMRG) method
\cite{white_prl_1992,white_prb_1993,white_prb_2005,schollwock_rmp_2005, schollwock_aop_2011,orus_aop_2014} with the matrix product state (MPS) ansatz \cite{schollwock_aop_2011, orus_aop_2014}. Specifically, we use a strictly single-site variant of DMRG (DMRG3S) with the subspace expansion \cite{Hubig_prb_2015}.
In our calculations the gauge fields are integrated out, and thus we do not need to use gauge-invariant tensor network \cite{tagliacozzo_prx_2014, buyens_prl_2014, silvi_njp_2014, kull_aop_2017}  but we can use standard globally symmetric $\mathbb{U}(1)$  MPS \cite{singh_pra_2010, singh_prb_2011}. This residual global symmetry corresponds to the conservation of the total dynamical charge $\sum_j \hat{Q}_j$.

The results from DMRG simulations, that we report, have been performed with maximum MPS bond dimension of $\chi=600$. To confirm the convergence of the DMRG sweeps, we continue the DMRG3S iterations until the energy difference in subsequent sweeps falls below $10^{-13}$. On the other hand, Fig.~\ref{fig:chi_convergence} shows the convergence of different quantities with respect to the bond dimension $\chi \leq 600$ for systems of sizes $L \leq 120$.
For example, the energy density converges close to the machine precision within $\chi \leq 500$ in the gapped regions -- confined and Higgs. On the other hand, as expected, the convergence is slower at the critical point due to the diverging correlation length. However, 
as shown in Fig.~\ref{fig:chi_convergence}\textbf{(b)}, the precision we attain at the critical point for $\chi = 500, 600$ is sufficient to perform the precise scaling analysis that we report in the main text.

\begin{figure}
\centering
\includegraphics[width=\linewidth]{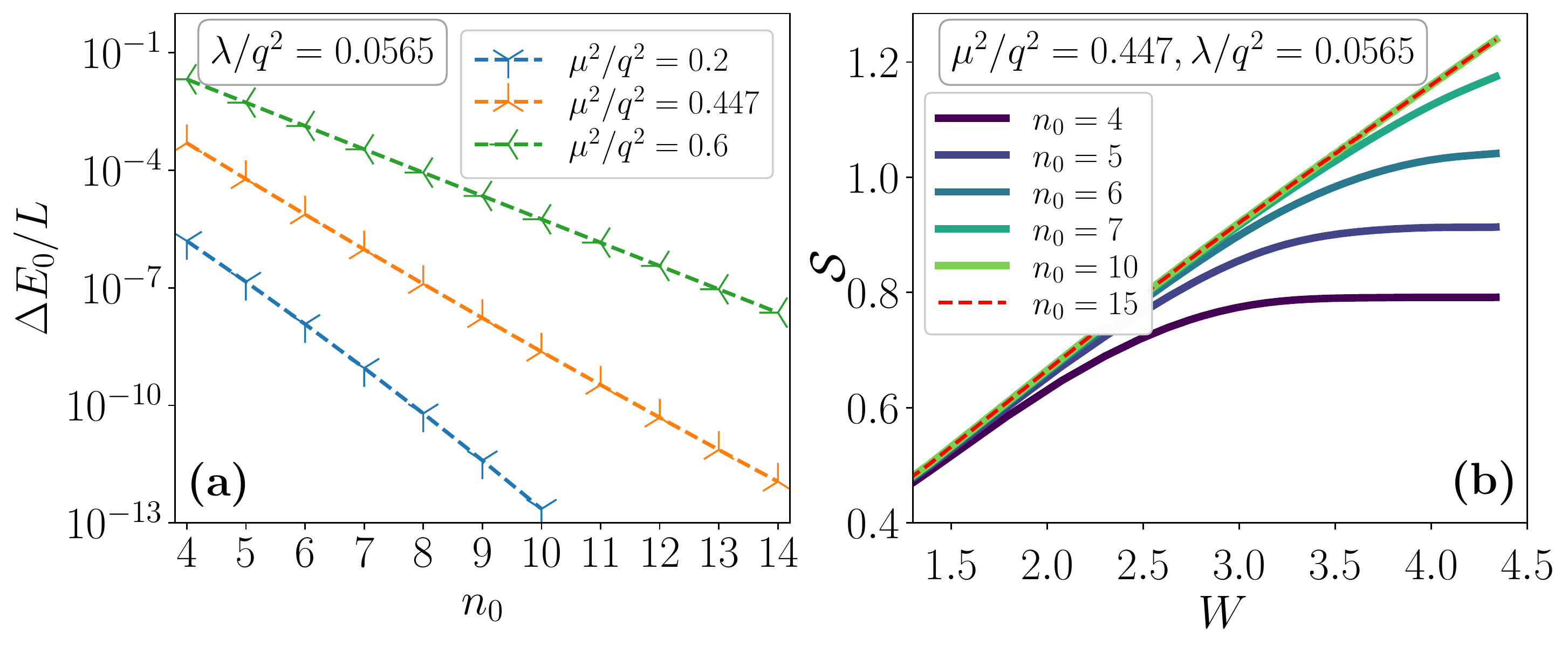}
\caption{\textbf{(a)} The convergence with respect to the maximum boson occupancy $n_0$ in three different points in the phase diagram, namely (1) a point in the confined region ($\mu^2/q^2 = 0.2$), (2)  the critical point ($\mu^2/q^2 = 0.447$), and (3) a point in the Higgs region ($\mu^2/q^2 = 0.6$) for a system of size $L=120$. Here we plot the relative errors in the energy density $\Delta E_0/ L$, where $\Delta E_0 = |E_0({n_0}) - E_0({n_0 +1})|$.
\textbf{(b)} The profile of the entanglement entropy $\mathcal{S}$ with respect to the cord length $W = \ln \left(\frac{2L}{\pi} \sin(\pi l/L) \right)$ at the critical point for $L \in [40, 120]$ and for different values of $n_0$. From the results presented it is clear that values of $n_0<10$ are not sufficient to capture the proper entropy scaling at the critical point.
\label{fig:n0_convergence}
}
\end{figure}	

Since, our system Hamiltonian does not conserve particle numbers for the individual bosonic species `$a$' and `$b$', but conserves their number difference, we verify the convergence with respect to the maximum bosonic occupancy $n_0$ for both the species. Specifically, we vary the bosonic cutoff $n_0$ in the range $[4, 15]$, and check for convergence of different observables (see Fig.~\ref{fig:n0_convergence}).
Clearly, as seen in Fig.~\ref{fig:n0_convergence}\textbf{(a)}, even a small $n_0 = 4$ is sufficient to faithfully capture the confined phase. However, this is not the case at the critical point or in the Higgs region. This is due to the fact that semiclassically the field $\phi$ attains non-zero expectation values
in the Higgs region. It is to be noted that in our calculations $\braket{\hat{\phi}}$ is trivially zero since it violates the global symmetry, while $\braket{\hat{\phi}^{\dagger} \hat{\phi}}$, and thereby individual $\braket{\hat{n}_a}$ and $\braket{\hat{n}_b}$, attain large expectation values in the Higgs region.
Moreover, Fig.~\ref{fig:n0_convergence}\textbf{(b)} shows that the smaller values of the bosonic cutoff $n_0$ are insufficient to capture the critical entropy scaling with respect to the cord length $W$ at the critical point. On the other hand, the entropy profile for $n_0=10$ is essentially identical to the one with $n_0=15$, capturing the proper entropy scaling with respect to the cord length.
Therefore, Fig.~\ref{fig:n0_convergence} clearly demonstrates that the results obtained with the bosonic cutoff $n_0=10$, provide accurate enough results, and this is the reason we have used this cutoff in our analysis. 

We do not exceed $n_0=10$ since the  local Hilbert space dimension is $d = (n_0+1)^2=121$, which is very large, and  the complexity of the DMRG3S algorithm scales as $\sim Ld\chi^3$. For this reasons 
we are forced to consider moderate  bond dimensions ($\chi \leq 600$) that also limit the range of system sizes ($L \leq 120$) that we can analyze accurately.

\begin{figure}
\centering
\includegraphics[width=\linewidth]{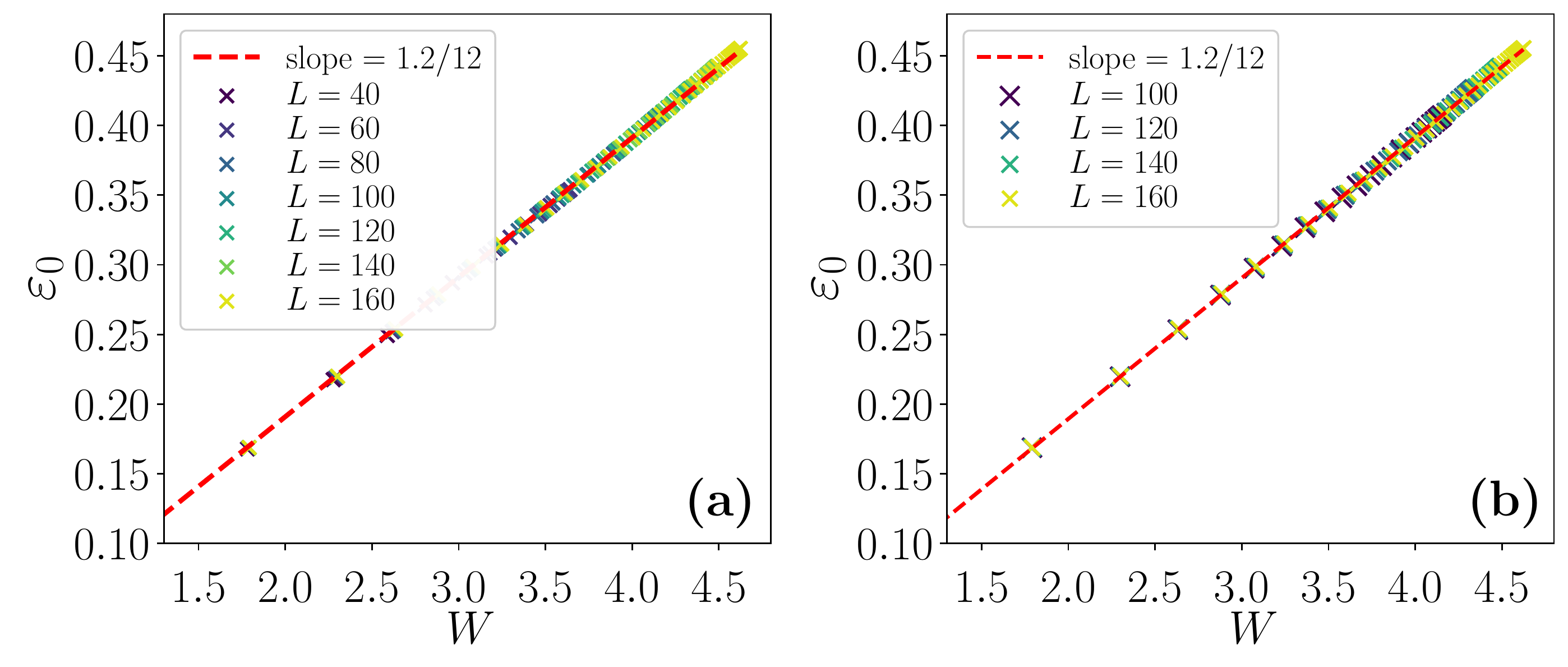}
\caption{The scaling of the entanglement ground state energy $\varepsilon_0$ with respect to the cord length $W$ for system-sizes \textbf{(a)} $L \in [40, 160]$ and \textbf{(b)} $L \in [100, 160]$.
\label{fig:entG160}
}
\end{figure}

On the other hand, Fig.~\ref{fig:chi_convergence}\textbf{(b)}  shows that the entanglement ground state energy $\varepsilon_0$ converges faster with respect to the bond dimension than the convergence in the entanglement entropy $\mathcal{S}$ (that requires the convergence of the full entanglement spectrum, not only of its ground state). Therefore, for this specific quantity, and just for a  comparison, we extend our finite size scaling analysis to the range $\varepsilon_0$ with $L \in [40, 160]$ keeping the bond dimension fixed at $\chi=600$. In Fig.~\ref{fig:entG160} we show the same scaling first in the whole range of $L \in [40, 160]$ and then for longer chains $L \in [100, 160]$. One could indeed think that the discrepancy between the central charge extracted from the entanglement entropy and the one extracted from the scaling of the first eigenvalue is ultimately a finite-size effect and that this would disappear in the thermodynamics limit. 

Unfortunately, in both the cases, the slopes are pretty robust to the choice of system sizes that we use in our fit, and remain fixed to $1.2/12$. Even using only the larger systems does not seem to induce a systematic  deviation towards the $1.5/12$ extracted from the scaling of the entanglement entropy reported in the main text. As a result, the level of accuracy, that we are able to achieve with our computational resources,
is extremely good and under control.
These extra analysis seems to  confirm the discrepancy between the value of the central charge extracted from the scaling of the entanglement ground state with respect to the one extracted from the scaling of the entanglement entropy.

\end{document}